\documentclass[reprint,superscriptaddress,twocolumn,secnumarabic,amssymb, nobibnotes, aps, prb]{revtex4-1}

\usepackage{chemformula} 
\usepackage[T1]{fontenc} 
\usepackage[ansinew]{inputenc}
\usepackage{gensymb}

\begin{document}

\title{
Electromagnetic dressing of the electron energy spectrum of Au(111) at high momenta
}

\author{Marius~Keunecke}
\email{mkeunec@gwdg.de}%
\affiliation{I. Physikalisches Institut, Georg-August-Universit\"at G\"ottingen, Friedrich-Hund-Platz 1, 37077 G\"ottingen, Germany}
\author{Marcel~Reutzel}
\email{marcel.reutzel@phys.uni-goettingen.de}%
\affiliation{I. Physikalisches Institut, Georg-August-Universit\"at G\"ottingen, Friedrich-Hund-Platz 1, 37077 G\"ottingen, Germany}

\author{David~Schmitt} %
\affiliation{I. Physikalisches Institut, Georg-August-Universit\"at G\"ottingen, Friedrich-Hund-Platz 1, 37077 G\"ottingen, Germany}

\author{Alexander~Osterkorn}%
\affiliation{Institut f\"ur Theoretische Physik, Georg-August-Universit\"at G\"ottingen, Friedrich-Hund-Platz 1, 37077 G\"ottingen, Germany}

\author{Tridev~A.~Mishra}%
\affiliation{Institut f\"ur Theoretische Physik, Georg-August-Universit\"at G\"ottingen, Friedrich-Hund-Platz 1, 37077 G\"ottingen, Germany}

\author{Christina~M\"oller}%
\affiliation{I. Physikalisches Institut, Georg-August-Universit\"at G\"ottingen, Friedrich-Hund-Platz 1, 37077 G\"ottingen, Germany}

\author{Wiebke~Bennecke}%
\affiliation{I. Physikalisches Institut, Georg-August-Universit\"at G\"ottingen, Friedrich-Hund-Platz 1, 37077 G\"ottingen, Germany}

\author{G.~S.~Matthijs~Jansen} %
\affiliation{I. Physikalisches Institut, Georg-August-Universit\"at G\"ottingen, Friedrich-Hund-Platz 1, 37077 G\"ottingen, Germany}

\author{Daniel~Steil} %
\affiliation{I. Physikalisches Institut, Georg-August-Universit\"at G\"ottingen, Friedrich-Hund-Platz 1, 37077 G\"ottingen, Germany}

\author{Salvatore~R.~Manmana}%
\affiliation{Institut f\"ur Theoretische Physik, Georg-August-Universit\"at G\"ottingen, Friedrich-Hund-Platz 1, 37077 G\"ottingen, Germany}

\author{Sabine~Steil} 
\affiliation{I. Physikalisches Institut, Georg-August-Universit\"at G\"ottingen, Friedrich-Hund-Platz 1, 37077 G\"ottingen, Germany}

\author{Stefan~Kehrein}%
\affiliation{Institut f\"ur Theoretische Physik, Georg-August-Universit\"at G\"ottingen, Friedrich-Hund-Platz 1, 37077 G\"ottingen, Germany}

\author{Stefan~Mathias}
\email{smathias@uni-goettingen.de}%
\affiliation{I. Physikalisches Institut, Georg-August-Universit\"at G\"ottingen, Friedrich-Hund-Platz 1, 37077 G\"ottingen, Germany}

\begin{abstract}
Light-engineering of quantum materials via electromagnetic dressing is considered an
on-demand approach for tailoring electronic band dispersions and even inducing topological phase transitions. For probing such dressed bands, photoemission spectroscopy is an ideal tool, and we employ here a novel 
experiment based on ultrafast photoemission momentum microscopy. Using this setup, we measure 
the in-plane momentum-dependent intensity fingerprints of the electromagnetically-dressed sidebands from a Au(111) surface for s- and p-polarized infrared driving. We find that at metal surfaces, due to screening of the driving laser, the contribution from Floquet-Bloch bands is negligible, and the dressed bands are dominated by the laser-assisted photoelectric effect. Also, we find that in contrast to general expectations, s-polarized light can 
dress free-electron states at large photoelectron momenta. Our results show that the dielectric response of the material 
must carefully be taken into account when using photoemission for the identification of light-engineered electronic band structures.
\end{abstract}

\maketitle
The on-demand femtosecond engineering of quantum materials by time-dependent external perturbations is a promising route for dynamical control of physical and chemical properties \cite{Basov17natmat}. For sufficiently strong external stimuli, the eigenstates of the equilibrium system are renormalized. The material's properties then depend on the crystal potential defined by the periodic arrangement of atoms in real-space, and, in addition, on the periodicity and strength of the external stimuli; novel phases of matter can be created as has been reviewed in the context of Floquet engineering~\cite{oka_floquet_2019, Rudner20nrp}. A particularly promising perturbation is the periodic electric field of an ultrashort laser pulse that can be used to engineer the energy-, momentum-, and time-dispersive band structure of a material \cite{Mucke01prl, Wang13sci, Sie14natmat, mahmood_selective_2016, Aeschlimann18phd, Reutzel20natcom, McIver20natphys, Lindner11natphys, sentef2015theory, Giovannini16nanolett, Liu19prl}. In a stroboscopic photoemission experiment, an intense driving field is used to build up an out-of-equilibrium band structure, while a weak probe field maps its current status~\cite{Wang13sci, mahmood_selective_2016, Aeschlimann18phd}.

\begin{figure}[hbt!]
    \centering
    \includegraphics[width=\linewidth]{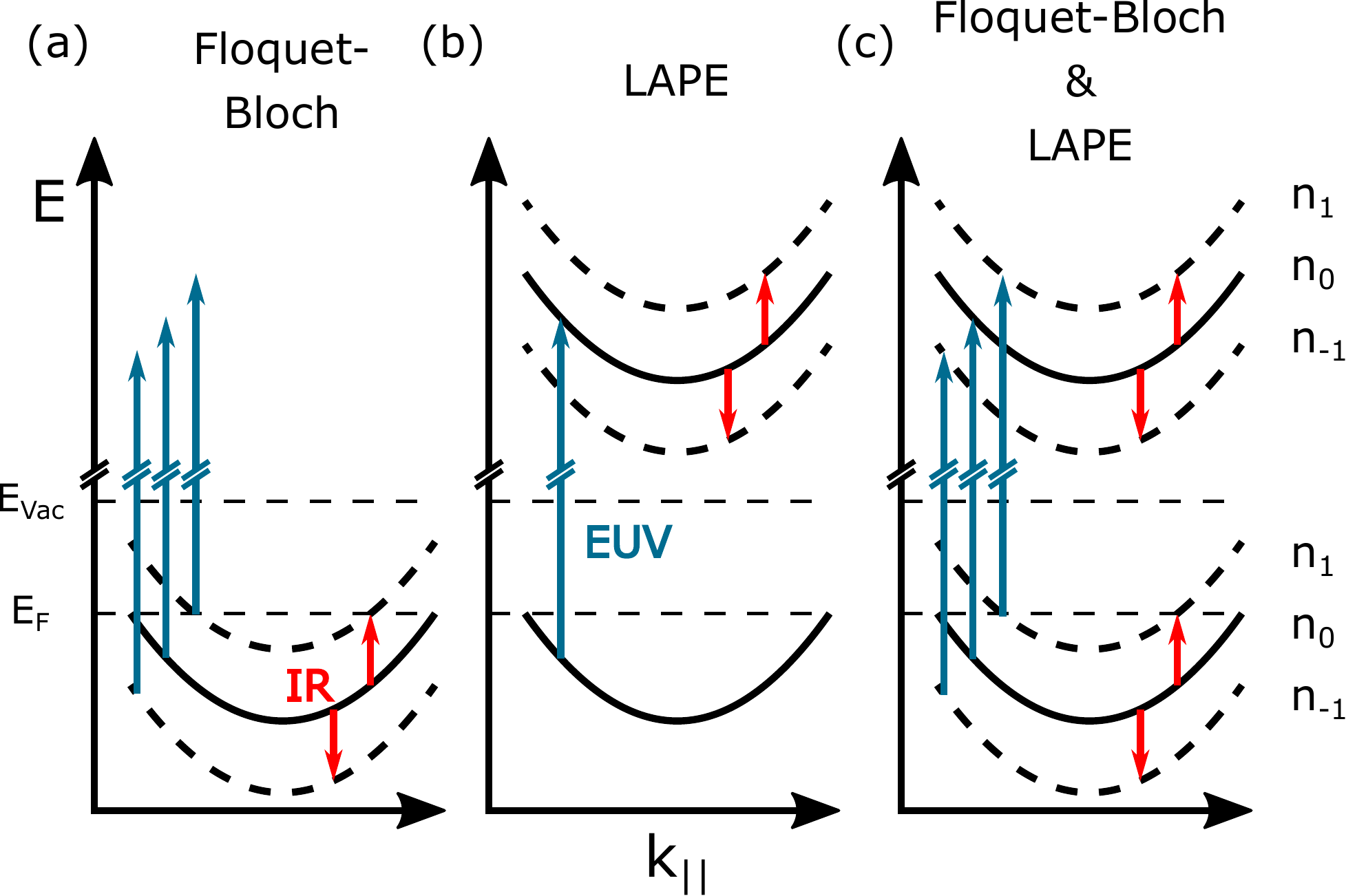}
    \caption{Schematics for the electromagnetic dressing  with IR light of (a) Bloch bands, yielding Floquet-Bloch bands, and (b) quasi-free electrons, leading to LAPE. In both scenarios, sidebands ($n_{\pm 1}$, dashed line) of the main photoemission spectral feature ($n_{0}$, solid line) are observed in the photoemission experiment. (c) Both processes terminate at the same final state energy, requiring the consideration of scattering amplitude between both processes.}
    \label{fig:floquet_lape_sketch}
\end{figure}

In such a time- and angle-resolved photoemission (TR-ARPES) experiment, the light-dressed band structure is evident in the formation of so-called Floquet-Bloch bands as sketched in Fig.~\ref{fig:floquet_lape_sketch}~(a). Those Floquet-Bloch bands appear as replicas of the main Bloch band spaced by the photon energy due to the time-periodicity of the driving field. Crucially, for the unambiguous identification of light-induced Floquet-Bloch bands in a two-color photoemission experiment, the laser-assisted photoelectric effect (LAPE) has to be considered in addition~\cite{miaja2006laser, saathoff_laser-assisted_2008}: The LAPE process creates sidebands of the main photoemission line at the same final state energy as would be expected for the photoexcitation of Floquet-Bloch bands [Fig.~\ref{fig:floquet_lape_sketch}~(b)]. However, while Floquet-Bloch bands represent a coherent modification of the electronic band structure of the material, LAPE is a final state effect, in which the photoemitted electron interacts with the electric field of the driving pulse in front of the surface. In consequence, LAPE does not have the potential to engineer material properties and is basically undesired in the quest of band structure engineering by light. Still, as both processes terminate at the same photoelectron energy, interference of both processes is expected~\cite{park_interference_2014} [Fig.~\ref{fig:floquet_lape_sketch}~(c)], which can be used to amplify the spectral signatures of Floquet-Bloch bands in TR-ARPES~\cite{park_interference_2014, mahmood_selective_2016}. 

In this letter, we study the contributions of Floquet-Bloch vs. LAPE bands from a Au(111) metal surface 
throughout the full accessible photoemission horizon, considering thus large in-plane momentum. 
Our analysis shows that on metal surfaces, the sideband formation is largely determined by LAPE. We further outline that not the impinging electric field strength of the driving light field builds up the sideband intensity, but the macroscopic screening response of the studied material defines the electric field strength available for dressing the electromagnetic energy spectrum, which can be crucially different for Floquet and LAPE that occur within and in front of the crystal, respectively.

\begin{figure}[hbt!]
    \centering
    \includegraphics[width=\linewidth]{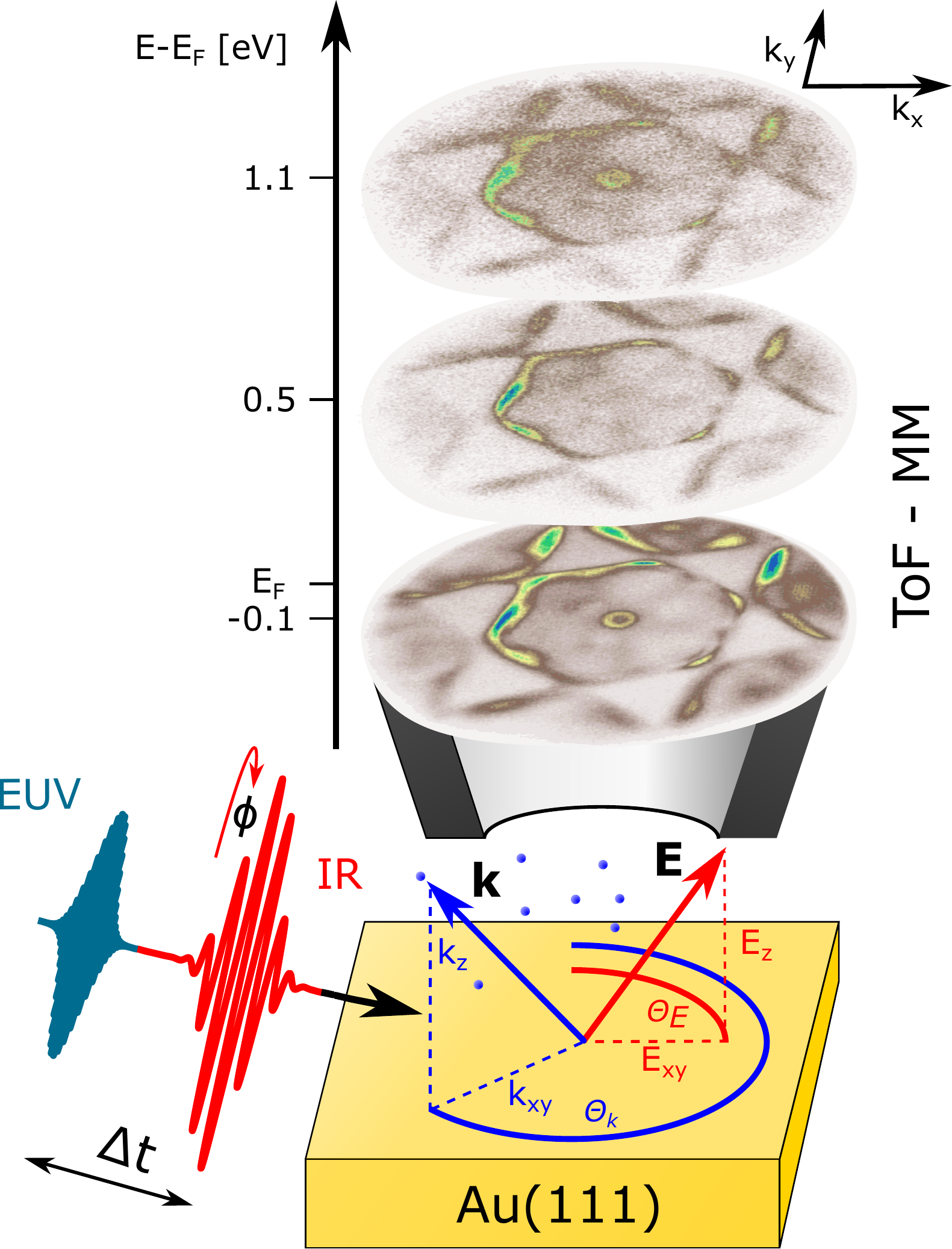}
    \caption{Time-resolved momentum microscopy experiment. In a stroboscopic experiment, we drive the Au(111) crystal with intense IR laser pulses and probe the instantaneous band structure with EUV light. The momentum microscope provides simultaneous access to the kinetic energy ($E_{\rm kin}$) and full in-plane momentum-resolved ($k_{x}$ and $k_y$) data sets. The accessible in-plane momentum range is limited by the photoemission horizon that scales with $k_{xy}\propto\sqrt{E_{\rm kin}}$. The experimental geometry is sketched in the bottom part. The drive and probe laser pulses impinge nearly collinear onto the surface at an angle of 22°. The coordinate frame of the in-plane electric field ($E_{xy}$, $\theta_E$) and momentum ($k_{xy}$, $\theta_k$) components is shown in polar coordinates; the out-of-plane components $E_z$ and $k_z$ are normal to the surface. The polarization of the driving light is tuned with a $\lambda/2$-plate and defined by the angle $\phi$.}
    \label{fig:setup}
\end{figure}

\begin{figure}[t!]
    \centering
    \includegraphics[width=\linewidth]{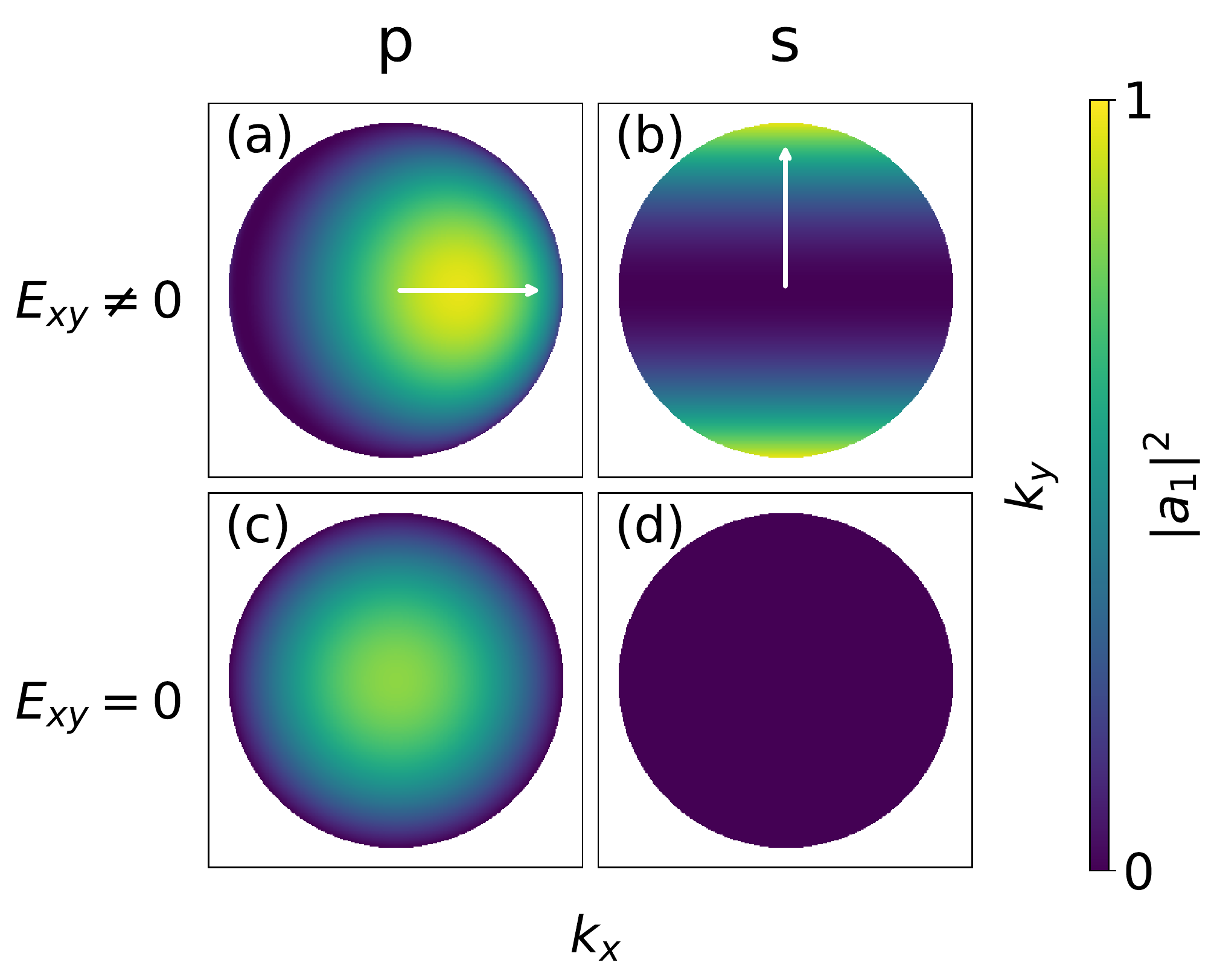}
    \caption{(a,b) Calculated in-plane momentum distributions of the LAPE sideband amplitude $|a_{1}|^2$ after Eq.~\ref{eq:alpha} for p- and s-polarized light impinging along $k_x$ in an oblique angle of incidence; the white arrow indicates the direction of the in-plane electric field component. Sideband yield can be expected for LAPE in both polarizations. (c,d) If the in-plane electric field components are screened ($E_{xy}=0$), no sidebands are expected for s-polarized light. In p-polarized driving, $|a_{1}|^2$ is symmetric around the $k_{xy}=0$. Note that all plots are visualized on the same color scale, in the full accessible photoemission horizon.}  
    \label{fig:sim}
\end{figure}

\begin{figure*}[hbt!]
    \centering
    \includegraphics[width=\linewidth]{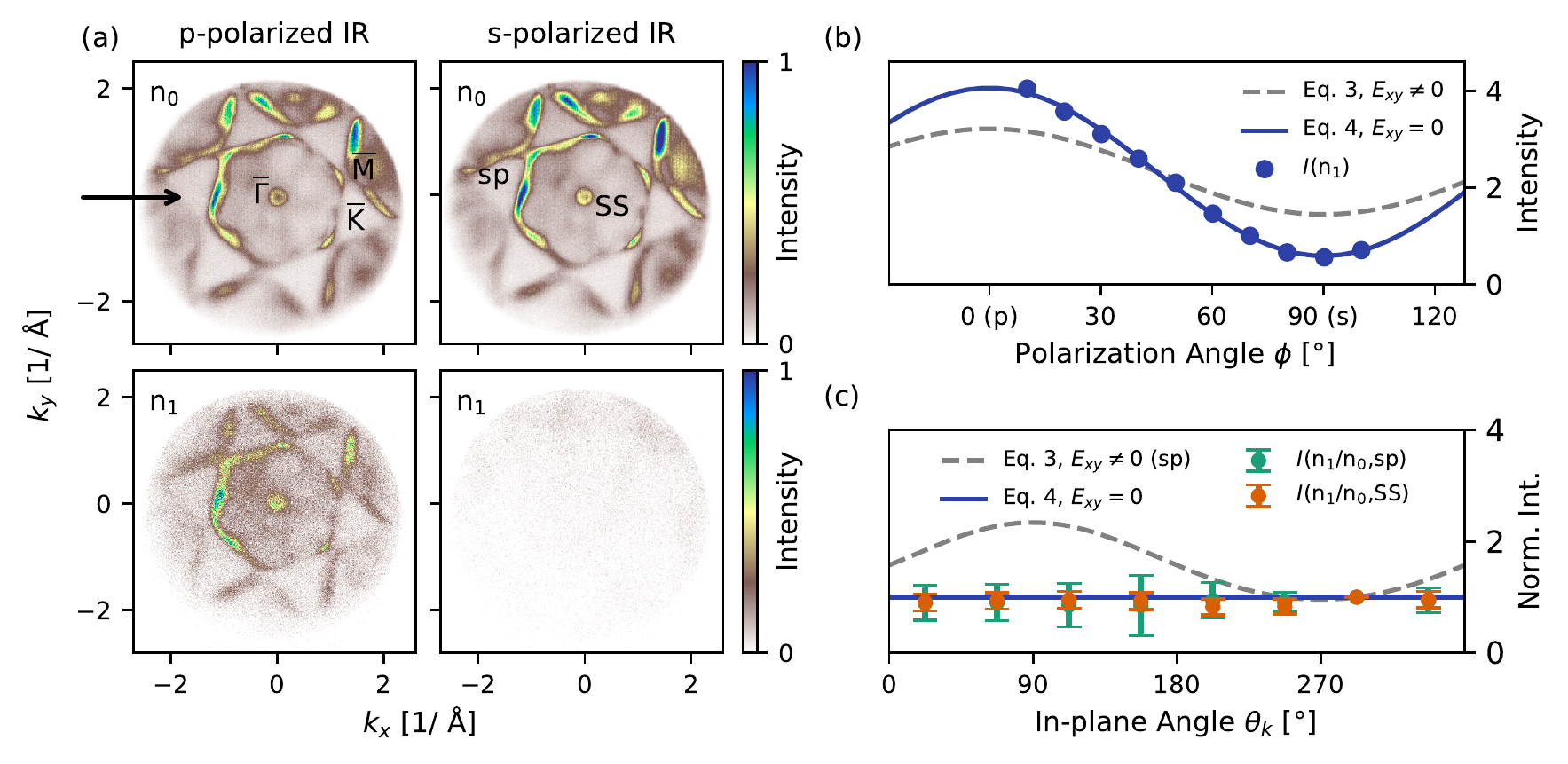}
    \caption{(a) ($k_x$, $k_y$)-resolved momentum maps extracted from the three-dimensional momentum microscopy data obtained with p-polarized EUV light and p- (left column) and s- (right column) polarized IR driving light in temporal overlap. The top row shows momentum maps taken close to the the Fermi-level that we label as the zero-photon-order sideband $n_0$. The high-symmetry points $\overline{\Gamma}$, $\mathrm{\overline{K}}$, $\mathrm{\overline{M}}$, and the Shockley surface band (SS) as well as the sp-band tranitions (sp) are indicated. The black arrow represents the direction of light incidence. The bottom row shows the first order sideband intensity ($n_1$) around $E-E_F$ = +1.1 eV above the Fermi-level for driving with p- (left) and s-polarized (right) IR light. (b) Systematic evaluation of the momentum integrated intensity of the first-order sideband $n_1$ when rotating from p- to s-polarized light. The data is well approximated with Eq.~\ref{eq:alphaExygleich0}, implicating that the in-plane electric field components are screened ($E_{xy}=0$). (c) Azimuthal dependence of the sideband intensity of the SS band and the sp-band transition (details on the data handling is provided in the supplemental material). The data can be fitted with Eq.~\ref{eq:alphaExygleich0}, indicating that the in-plane component of the electric field is efficiently screened.
    }  
    \label{fig:ARPES}
\end{figure*}

In Fig.~\ref{fig:setup}, we show exemplary TR-ARPES data obtained on the pristine Au(111) surface, which is driven with an infrared laser pulse  at an incidence fluence of 5~mJ/cm\textsuperscript{2} (IR, $\hbar\omega=1.2$~eV, nearly p-polarized, $\approx40$~fs). Photoemission is induced by an extreme ultraviolet (EUV) light pulse generated by a table-top high-harmonic generation beamline operated at 1~MHz repetition rate ($\hbar\omega=26.5$~eV, p-polarization)~\cite{Keunecke20timeresolved}. With our novel time-of-flight momentum microscope (ToF-MM)~\cite{medjanik_direct_2017}, we measure the photoelectron yield as a function of kinetic energy ($E_{\rm kin}$) and full in-plane momentum ($k_{x}$ and $k_y$)  \cite{Keunecke20timeresolved, jansen_efficient_2020}. The EUV photons map the Fermi level to a kinetic energy of 21.2~eV (photon energy $\hbar\omega=26.5$~eV minus vacuum level $E_V = 5.3$~eV), facilitating the access of in-plane momenta up to $k_{x,y}\approx$ $2.4$ \AA$^{-1}$, thus covering the full first surface Brillouin zone and parts of the second Brillouin zone of Au(111). Photoelectrons that are detected at higher kinetic energies (corresponding to $E-E_F >$ 0~eV in Fig. 2) must have interacted with the EUV probe as well as the IR driving laser field before their experimental detection. This two-color interaction can be of multiple origin: (i) In real interband transitions, electrons can be excited by the IR pulse into initially unoccupied bands and subsequently photoexcited above the vacuum level with the EUV pulse. (ii) EUV photons might probe the Floquet-Bloch bands of positive photon order that are built up by the intense IR laser field. Finally, (iii) quasi-free photoelectrons emitted by the EUV pulse might be dressed by the IR laser field in front of the surface, evident as LAPE. However, for our experimental parameters, we do not map any real unoccupied bands that could be photoexcited via process (i), even so this has been shown for Au(111) when using photon energies in the infrared regime \cite{Yan15natcom,Reutzel20prb}. Instead, the photoemission spectral replica features in Fig.~\ref{fig:setup} can be attributed to first order sidebands $n_1$ with the momentum resolved intensity distribution $I_1\left(k_{xy}, \theta_k,k_{z}\right)$ of the zero-photon-order photoemission spectral feature $n_0$ ($I_0\left(k_{xy}, \theta_k,k_{z}\right)$). In the following, we discuss the in-plane momentum resolved intensity distributions of these sidebands that are generated by processes (ii) and (iii). Especially, we will focus on the formation of sidebands at large in-plane momenta ($k_{xy}\geq 1 $\AA$^{-1}$), which has to our knowledge not been studied previously.


Before we go into detail of our experimental results, we first calculate the expected momentum fingerprints of the first order sidebands; their photoemission yield scales with
\begin{equation}
I_1\left(k_{xy}, \theta_k,k_{z}\right)\sim I'_0\left(k_{xy}, \theta_k,k_{z}\right)\times|a_{1}|^2,
\label{eq:yield}
\end{equation} 
where $I'_0\left(k_{xy},\theta_k ,k_{z}\right)$ is the photoemission yield of the undriven system, and $|a_1|^2$ is the sideband amplitude. In early work by Miaja-Avila \emph{et al.}~\cite{miaja2006laser,saathoff_laser-assisted_2008}, sidebands in two-color TR-ARPES experiments have been explained by pure LAPE physics. Only later, Gedik and coworkers~\cite{Wang13sci,mahmood_selective_2016} reasoned that simultaneously occurring Floquet engineering might be observed in these experiments. Thus, in general, the sideband amplitude $|a_1|^2$ can contain contributions from Floquet and LAPE processes that we describe with the $\beta$ and $\alpha$ parameters, respectively. Following the notation of Park~\cite{park_interference_2014}, the overall sideband amplitude is then given with
\begin{equation}
|a_1|^2 \sim \frac{1}{4} \left( \beta - \alpha \right)^2,
\label{eq:alphabeta}
\end{equation}
which intrinsically contains scattering amplitude between both processes. For LAPE, electromagnetic dressing occurs in the photoemission continuum; the free electron final states can be described by Volkov states. In an electron scattering description \cite{madsen2005strong, baggesen2008theory, park_interference_2014}, the LAPE parameter of the first-order sideband can be calculated with
\begin{equation}
\alpha \sim \left(\frac{e}{m_e\omega_{IR}^2}(E_{xy}k_{xy}\cos(\theta_{k}-\theta_{E})+E_{z}k_{z})\right),
\label{eq:alpha}
\end{equation}
as we detail in the supplemental material; the in-plane electric field and momentum is written in polar coordinates, as labelled in Fig.~\ref{fig:setup}. $\omega_{IR}$ is the driving light frequency, and $m_e$ and $e$ the electron mass and charge.

Before considering the contribution of the Floquet parameter $\beta$ to the sideband amplitude, we first calculate the expected momentum fingerprints of LAPE sidebands for p- and s-polarized driving light based on Eq.~\ref{eq:alphabeta} ($\beta=0$) and \ref{eq:alpha} [Fig.~\ref{fig:sim} (a, b)]. Clearly, the in-plane momentum resolved sideband amplitude $|a_1|^2$ shows distinct azimuthal asymmetries that become more prominent with increasing $k_{xy}$, where the contribution of the term $\approx E_{xy}k_{xy}$ in Eq.~\ref{eq:alpha} becomes comparable to the out-of-plane component $E_z k_z$. Importantly, we can directly see that LAPE sidebands should also appear in the case of s-polarized driving. This stands in contrast to typical expectations in experiments that are performed close to the $\overline{\Gamma}$-point, where especially LAPE sidebands are not expected as $k_{xy}$ is considered negligibly small~\cite{mahmood_selective_2016}.


Based on these momentum fingerprints originating from LAPE, we turn back to the experimental results obtained on Au(111). Fig.~\ref{fig:ARPES} shows selected ($k_x$, $k_y$) momentum maps of the three-dimensional momentum microscopy data sets. Centered at the $\overline{\Gamma}$-point, the occupied part of the SS band is resolved for energies close to the Fermi level (top row). At larger $k_{x}$ and $k_y$, the full hexagonal structure of the sp-band transition within the first and the second surface Brillouin zone are resolved. In the two-color experiment with p-polarized driving light, replica sideband structures of the SS band and the sp-band 1.2~eV above the original structures are seen (Fig.~\ref{fig:ARPES}~(a), bottom left). In the following, we will further evaluate the sideband intensities throughout the full measured 3D data set.

First, we systematically vary the polarization angle $\Phi$ of the driving laser field from p- to s-polarization, keeping all other parameters fixed. The in-plane momentum-integrated photoemission yield of the first order sideband, I$_1$($k_{xy},\theta_k, E-E_F \approx +1.1$ eV), is shown in Fig.~\ref{fig:ARPES}~(b). The intensity of the IR driving-induced sideband features drops systematically when the out-of-plane field component is reduced by rotating to overall s-polarization. The associated momentum maps for p- and s-polarized driving light are shown in Fig.~\ref{fig:ARPES}~(a), bottom left and right. Strikingly, no distinct photoemission spectral features of the sidebands are resolved for s-polarized driving light within our noise level (which is slightly increased due to photoemission with residual light of a neighboring harmonic with $\hbar\omega=31.4$~eV, see Ref. \cite{Keunecke20timeresolved} and supplemental material). Based on the calculations shown in Fig.~\ref{fig:sim}, this observation is unexpected at first: For $k_{xy}>2$ \AA$^{-1}$ and $\Theta_k=0$\degree and $180$\degree, i.e. close to the edge of the photoemission horizon and perpendicular to the plane of incidence of the driving light, sideband intensitites should be resolved.

This observation can be understood, however, when taking screening of the IR electromagnetic field in front of the high-electron density Au(111) crystal into account. At the metallic surface, in-plane electric field components with driving frequencies below the plasmon frequency are reflected with near unity; the local in-plane electric field strength within the crystal and in front of the surface is close to zero due to destructive interference of the incoming and outgoing electric field. Thus, at the metallic surface, Eq.~\ref{eq:alpha} must be reconsidered with $E_{xy}=0$ to
\begin{equation}
\alpha \sim \frac{e}{m_e\omega_{IR}^2}E_{z}k_{z}.
\label{eq:alphaExygleich0}
\end{equation}
In the bottom row of Fig.~\ref{fig:sim}, the revised calculations using Eq.~\ref{eq:alphaExygleich0} are shown. In agreement with experiment, no sideband amplitude is present in s-polarized driving. Following this reasoning, we fit the polarization dependent photoemission yield in Fig.~\ref{fig:ARPES}~(b) with Eq.~\ref{eq:alphaExygleich0}, which nicely describes the experimental results (blue fit). In contrast, if we include in-plane field components, i.e. use Eq.~\ref{eq:alpha}, the data is not described to a satisfactory level (grey dashed line).


Having identified the absence of sidebands caused by in-plane field components of the driving laser, we now turn our attention again to p-polarized driving light that contains both in-plane and out-of-plane field components. Since the surface normal component of the electric field ($E_z$) is not screened in front of the surface, LAPE sidebands must be expected. Indeed we clearly observe sidebands (Fig.~\ref{fig:ARPES}~(a), bottom left), and now analyze our experimental data to verify either the asymmetric or symmetric intensity fingerprint as shown in Fig.~\ref{fig:sim}~(a) or ~\ref{fig:sim}~(c). Therefore, we plot in Fig.~\ref{fig:ARPES}~(c) the measured relative sideband intensity $I(n_1)/I(n_0)$ of the SS band and the sp transition as a function of azimuthal angle $\Theta_k$. For both cases, $I(n_1)/I(n_0)$ is not modulated with $\Theta_k$ (for analysis details see supplemental material). For the SS band located at the $\overline{\Gamma}$-point, this is expected as $k_{xy}\approx0.15$ \AA$^{-1}$\hspace{.05cm} and thus negligible small; Eq.~\ref{eq:alpha} and Eq.~\ref{eq:alphaExygleich0} would yield similar sideband intensities even with contributions from $E_{xy}\neq0$. However, the sp band transition is probed at $k_{xy}\approx1$ \AA$^{-1}$ \hspace{.05cm}, but is still not modulated with $\theta_k$. Thus, as well under p-polarized driving, we can identify the screening of the in-plane electric field components as they do not contribute to the electromagnetic dressing of the energy spectrum of the metallic surface at high momenta.


Up to now, we have identified two major conclusions: (a) Driving with s-polarized light should create LAPE sidebands with increasing intensity towards the photoemission horizon [see Fig.~\ref{fig:sim}~(b)]. (b) If the studied material system is highly reflective for the applied driving frequency, like in our case Au(111) for IR light, the in-plane electric field components are effectively screened in the bulk and in front of the surface; only surface normal field components can lead to the formation of sidebands. Concerning (a), this observation is rather crucial when searching for light-engineered band structures at the edges of the surface Brillouin zone, for example, on graphene and other two-dimensional materials \cite{sentef2015theory,Giovannini16nanolett}: also in case of s-polarized excitation, contributions of LAPE have to be considered.


The open question is whether we can identify Floquet-Bloch contributions to the measured sideband yield in the two-color photoemission data obtained on Au(111). Therefore, it is insightful to calculate $\beta$ and thus the expected momentum fingerprint of the Floquet sideband amplitude. In the supplemental material, we approximate the form of $\beta$, which depends on the initial state momentum dispersion, for the two-dimensional, parabolic surface band (the SS band) and a bulk band transition with a more complex dispersion relation (the sp band transition). In addition, we consider the contribution of the interference term $(2\beta\alpha)$ to $|a_1|^2$ (cf. Eq.~\ref{eq:alphabeta}), and find an interesting result: in the case of perfectly parabolic bands, interference between Floquet-Bloch and LAPE bands can induce complete destructive interference, i.e. no sidebands would be observable in photoemission.

However, in the case of Au(111), we argue that the Floquet contribution is in any case negligible, and the measured photoemission yield of the sidebands in Fig.~\ref{fig:ARPES}~(a) is caused by LAPE electrons only, as can be understood by considering screening of the driving electromagnetic field at the metallic surface. First, when considering the in-plane field components, like discussed above for LAPE, no Floquet-Bloch sideband amplitude can be expected as $E_{xy}\approx0$ due to screening. Importantly, this statement is independent of the explicit form of the $\beta$ parameter, and thus true for all initial states, independent on their momentum dispersion. Second, for electromagnetic dressing with the out-of-plane field component ($E_z$), the situation is slightly more complex. Both, in front of the surface and in the bulk material, $E_z$ is finite and can thus couple to the $k_z$ component of the initial (Floquet) and the final state (LAPE); the relative strength of the Floquet contribution will then depend on the explicit dispersion of the initial state. Here, we neglect the contribution of the initial state dispersion and only estimate the relative strength of $E_z$ as follows: LAPE is considered to occur close to the crystal, where the surface can act as a sink for momentum conservation in the light-dressing process. In contrast, Floquet-Bloch bands would be created within the bulk material; the electric field has to penetrate into the crystal. In the simplest approach, considering an abrupt metal-vacuum interface that could be described via Fresnel equations, the surface normal field component discontinuously drops at the surface barrier. With the dielectric function of gold \cite{Olmon12prb} and 1.2~eV driving light, we estimate that $E_z$ in the bulk material drops to 2\% of its value at the surface. This estimation clearly illustrates that the measured sideband yield in p-polarized driving is dominated by LAPE physics. In addition, it exemplifies how critical the screening capabilities of the material have to be considered, if photoemission band mapping is the method of choice for the investigation and identification of light-engineered electronic band structures. We want to emphasize, however, that a light-induced coherent manipulation of the electronic band structure from a metal surface is possible and also has recently been observed using interferometric photoemission techniques~\cite{Reutzel20natcom}.


In conclusion, we present a systematic evaluation of the electromagnetic dressing of the electron energy spectrum at high in-plane momenta, i.e. within the full measured photoemission horizon. In contrast to photoemission experiments focusing on features close to the $\overline{\Gamma}$-point ($k_{xy}\approx0$ \AA$^{-1}$) \cite{miaja2006laser, saathoff_laser-assisted_2008,mahmood_selective_2016}, for $k_{xy}$ near the photoemission horizon, the in-plane electric field components $E_{xy}$ can, in principle, induce light-dressing of free electron states, i.e. LAPE. However, our analysis shows that not the external electric field strength defines the dressing response, but that the local electric field strength at the crystal has to be considered. Thus, depending on the frequency dependent dielectric tensor, sideband yield can be largely suppressed (and potentially enhanced) in the two-color photoemission experiment.

Our analysis further shows that the distinct separation of Floquet-Bloch and LAPE contributions in a two-color photoemission experiment is challenging. We believe that from modelling of the expected momentum fingerprints for bands with specific initial state momentum dispersions, such as done for two-dimensional linear bands in Ref.~\cite{park_interference_2014,mahmood_selective_2016}, and carried out for parabolic bands in the supplemental material, further insight can be gained. Here, especially for the case of 3D dispersive bulk bands, further theoretical work is needed.

Beyond the macroscopic material properties that define the local electric field strength that can potentially build up Floquet-Bloch (and LAPE) sidebands, further theory efforts suggest that also the time scale of decoherence of the optical excitation \cite{Kandelaki18prl,Sato20arxiv,Schueler20arxiv}, and the pulse duration of the driving field in relation to the optical cycle duration \cite{Kalthoff18prb,Giovannini16nanolett} can hinder the creation of light-engineered band structures, even so sufficient electric field strength is available for efficient dressing. Based on our results and those predictions, we speculate that for the on-demand creation and detection of light-engineered band structures, one thus first has to consider the macroscopic material properties, and second choose driving conditions that guarantee a minimum phase space into which energy can dissipate.

\begin{acknowledgements}

This work was funded by the Deutsche Forschungsgemeinschaft (DFG, German Research Foundation) - 217133147/SFB 1073, projects B07 and B03. M.R. and G.S.M.J. acknowledge funding by the Alexander von Humboldt Foundation. S.S. acknowledges the Dorothea Schl\"ozer Postdoctoral Program for Women.

\end{acknowledgements}

\end{document}


\title{
Supplemental Information:\\
Electromagnetic dressing of the electron energy spectrum of Au(111) at high momenta
}

\author{Marius~Keunecke}
\email{mkeunec@gwdg.de}%
\affiliation{I. Physikalisches Institut, Georg-August-Universit\"at G\"ottingen, Friedrich-Hund-Platz 1, 37077 G\"ottingen, Germany}
\author{Marcel~Reutzel}
\email{marcel.reutzel@phys.uni-goettingen.de}%
\affiliation{I. Physikalisches Institut, Georg-August-Universit\"at G\"ottingen, Friedrich-Hund-Platz 1, 37077 G\"ottingen, Germany}

\author{David~Schmitt} %
\affiliation{I. Physikalisches Institut, Georg-August-Universit\"at G\"ottingen, Friedrich-Hund-Platz 1, 37077 G\"ottingen, Germany}

\author{Alexander~Osterkorn}%
\affiliation{Institut f\"ur Theoretische Physik, Georg-August-Universit\"at G\"ottingen, Friedrich-Hund-Platz 1, 37077 G\"ottingen, Germany}

\author{Tridev~A.~Mishra}%
\affiliation{Institut f\"ur Theoretische Physik, Georg-August-Universit\"at G\"ottingen, Friedrich-Hund-Platz 1, 37077 G\"ottingen, Germany}

\author{Christina~M\"oller}%
\affiliation{I. Physikalisches Institut, Georg-August-Universit\"at G\"ottingen, Friedrich-Hund-Platz 1, 37077 G\"ottingen, Germany}

\author{Wiebke~Bennecke}%
\affiliation{I. Physikalisches Institut, Georg-August-Universit\"at G\"ottingen, Friedrich-Hund-Platz 1, 37077 G\"ottingen, Germany}

\author{G.~S.~Matthijs~Jansen} %
\affiliation{I. Physikalisches Institut, Georg-August-Universit\"at G\"ottingen, Friedrich-Hund-Platz 1, 37077 G\"ottingen, Germany}

\author{Daniel~Steil} %
\affiliation{I. Physikalisches Institut, Georg-August-Universit\"at G\"ottingen, Friedrich-Hund-Platz 1, 37077 G\"ottingen, Germany}

\author{Salvatore~R.~Manmana}%
\affiliation{Institut f\"ur Theoretische Physik, Georg-August-Universit\"at G\"ottingen, Friedrich-Hund-Platz 1, 37077 G\"ottingen, Germany}

\author{Sabine~Steil} 
\affiliation{I. Physikalisches Institut, Georg-August-Universit\"at G\"ottingen, Friedrich-Hund-Platz 1, 37077 G\"ottingen, Germany}

\author{Stefan~Kehrein}%
\affiliation{Institut f\"ur Theoretische Physik, Georg-August-Universit\"at G\"ottingen, Friedrich-Hund-Platz 1, 37077 G\"ottingen, Germany}

\author{Stefan~Mathias}
\email{smathias@uni-goettingen.de}%
\affiliation{I. Physikalisches Institut, Georg-August-Universit\"at G\"ottingen, Friedrich-Hund-Platz 1, 37077 G\"ottingen, Germany}

\maketitle

\section{Experimental details}

\subsection{Experimental setup}

The momentum microscope and the 1~MHz HHG beamline are described in detail in Ref.~\cite{Keunecke20timeresolved}. Briefly, we use the compressed and frequency doubled output of a fiber-based amplifier system (Active Fiber Systems) to generate high harmonics in an Ar gas jet. The p-polarized 11\textsuperscript{th} ($\hbar\omega=26.5$~eV) harmonic is separated via two multilayer mirrors and focused onto the Au(111) crystal that is mounted on a hexapod in the momentum microscope. A part of the output of the amplifier system is separated and used as the driving light to electromagnetically dress the electron energy spectrum of Au(111).  A $\lambda/2$ waveplate and an attenuator allow for manipulation of the polarization and the pulse energy, respectively, before the pulses are focused onto the sample nearly collinear with the 11\textsuperscript{th} harmonic. A delay stage ensures control over the temporal delay between the driving and the probing laser pulse.

The time-of-flight based momentum microscope provides simultaneous access to the photoelectron´s kinetic energy ($E_{\mathrm{Kin}}$) and its in-plane momentum ($k_x$ and $k_y$). Details on the working principle of the microscope are provided in Ref.~\cite{Keunecke20timeresolved, kromker_development_2008, medjanik_direct_2017,kutnyakhov_time-_2020}. The measurements reported in the main text are performed with the following parameters:
(i) The extractor voltage of the microscope is set to 12 kV facilitating access to photoemission angles of up to $\pm 90$\degree, i.e. the full photoemission horizon. The extractor voltage is held sufficiently small to minimize effects of field emission. (ii) In the time-of-flight drift tube, the drift voltage is set to 40~V that defines the energy resolution of the momentum microscope (see below) (iii) The read-out of the time- and position sensitive delay-line detector [i.e. the photoelectron counts I($E$,$k_{x}$,$k_{y}$)] are stored for each measurement in the .tif format; metadata like laser power, delay, microscope lens voltages, and driving light polarization are stored separately for each measurement run. 

The Au(111) crystal is cleaned by subsequent heating (650~K) and Ar-sputtering cycles prior to the measurement. The sample quality is checked by the well-resolved Shockley surface band of the Au(111) surface. All measurements have been performed at room temperature.

\begin{figure*}[hbt!]
    \centering
    \includegraphics[width=\linewidth]{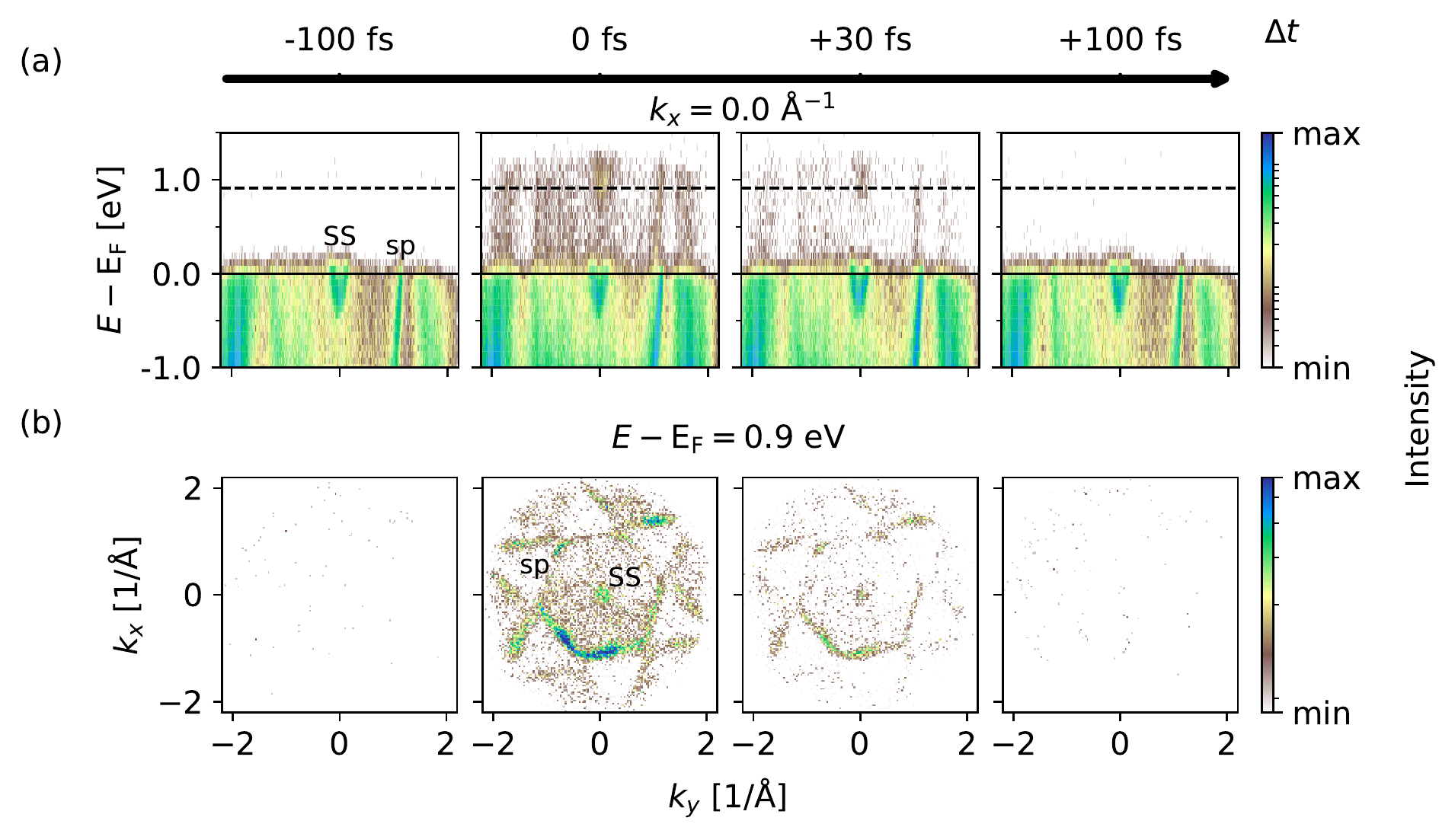}
    \caption{Temporal evolution of the sideband yield for selected $\Delta t$. (a) The ($E$, $k_y$)-resolved spectra are retrieved by slicing the experimentally obtained three-dimensional data set in $k_{y}$ direction and integrating over a 0.13~\AA$^{-1}$ large region in $k_{x}$ direction. The dashed and solid lines depict the momentum cut shown in (b) and the Fermi level, respectively. (b) Likewise, the data can be plotted in a ($k_x$, $k_y$)-resolved manner. The integration time for each $\Delta t$ is 15~min. Sideband yield is maximum in temporal overlap and follows the electric field strength of the driving laser field.}  
    \label{fig:timetrace}
\end{figure*}

\begin{figure*}[hbt!]
    \centering
    \includegraphics[width=\linewidth]{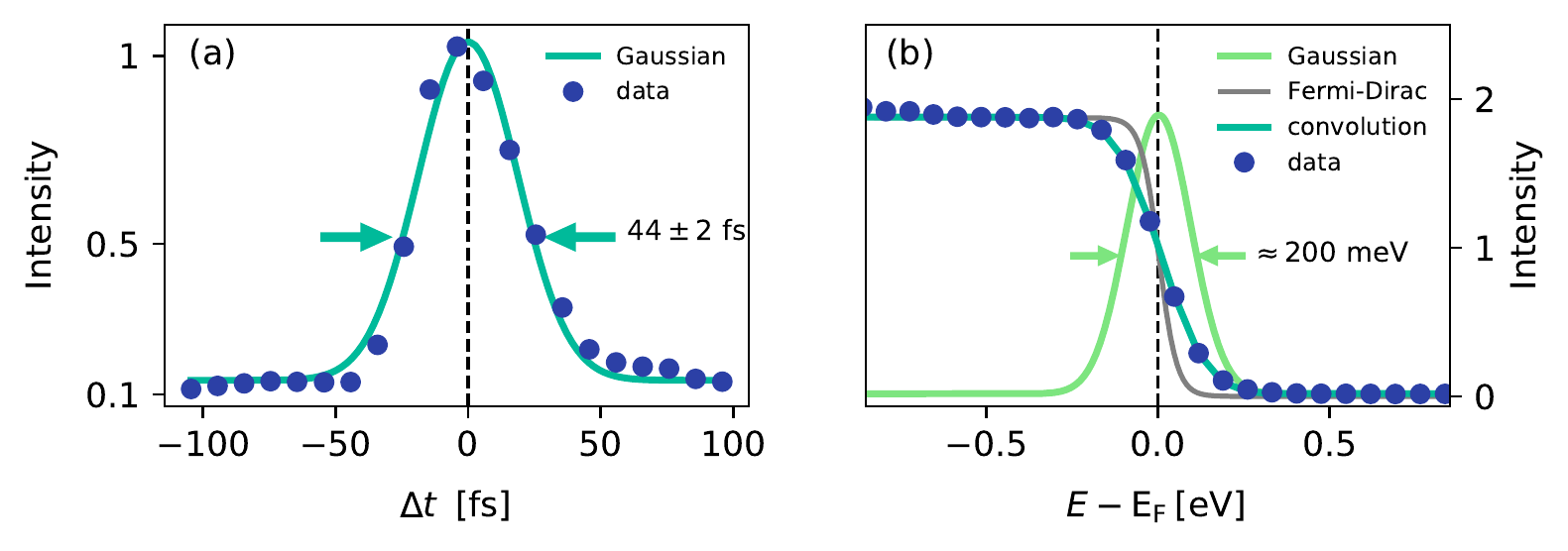}
    \caption{Time- and energy-resolution of the time-resolved momentum microscopy experiment. (a) The time-resolution is extracted in a cross correlation experiment by considering the photoemission yield of the first-order sideband following instantaneously the applied electric field strength of the driving laser. A Gaussian fit to the data yields a FWHM of $44\pm 2$~fs. (b) The energy resolution is extracted by fitting the convolution of a Fermi-Dirac distribution (300~K) and a Gaussian broadening to the measured Fermi edge of Au(111) (for the $\Delta t= -100$~fs case). The FWHM of the Gaussian broadening is in the range of $\approx 200$~meV. Details on the time- and energy-resolution of the experiment are discussed in Ref.~\cite{Keunecke20timeresolved}.}  
    \label{fig:reso}
\end{figure*}

\subsection{Energy- and time-resolution}

Our setup facilitates the real-time measurement of the temporal evolution of the ultrafast charge carrier dynamics with full energy and in-plane momentum resolution~\cite{Keunecke20timeresolved}. This is evident in Fig.~\ref{fig:timetrace} where we show exemplary ($E$, $k_y$) and ($k_x$, $k_y$) cuts for various delays ($\Delta t$) of the driving IR and the probing EUV pulse. By scanning the delay, we find $\Delta t=0$~fs at the position where maximum sideband intensity is observed, similar as done in Ref.~\cite{Keunecke20timeresolved, saathoff_laser-assisted_2008}. From this data, we estimate the temporal and the energetic resolution of our experiment to $44\pm2$~fs (FWHM of the cross-correlation) and $\approx200$~meV, respectively (Fig.~\ref{fig:reso}). To obtain the data in Fig.~4 of the main text, for each polarization of the driving light field, photoelectron counts are integrated for 2~h.

\subsection{Contributions to the photoelectron background signal}
\label{sec:background}

The HHG beamline is optimized for EUV light with $\hbar\omega_{\rm EUV}=26.5$~eV (11$^{\rm th}$-harmonic). However, as detailed in Ref.~\cite{Keunecke20timeresolved}, the extinction ratio to the neighboring 13$^{\rm th}$-harmonic is estimated to ~1:470. In consequence, spectral contributions of the 13$^{\rm th}$-harmonic have to be carefully considered in the evaluations presented in Fig.~4 of the main text. 

\begin{figure}[hbt!]
    \centering
    \includegraphics[width=0.6\linewidth]{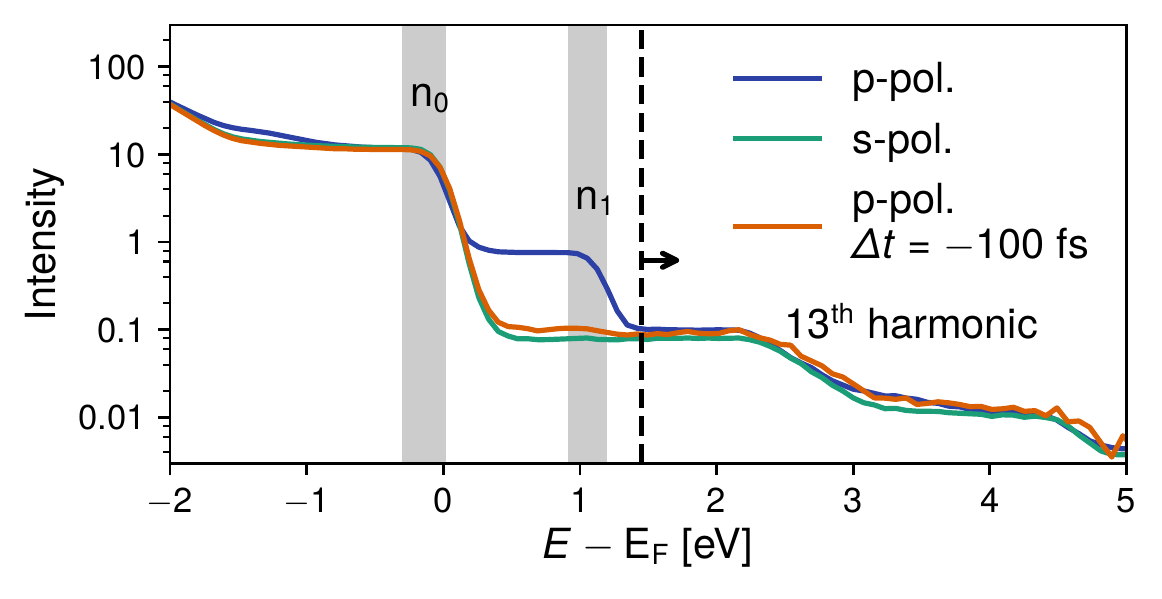}
    \caption{Momentum integrated energy distribution spectra for p- and s-polarized driving laser pulses; in addition, an EDC is shown for $\Delta t = -100$ fs (scaled) where no sidebands are expected. The integration windows for the zero- and first-order sidebands used in Fig.~4 of the main text are indicated by grey vertical areas. For, $E-\mathrm{E_{F}}>1.2$~eV, major photoelectron counts are induced by linear photoemission with the 13\textsuperscript{th} harmonic.}  
    \label{fig:spectrum}
\end{figure}

In Fig.~\ref{fig:spectrum}, energy distribution curves (EDCs) are shown for the nearly p- and s-polarized driving light. In addition, an EDC obtained away from temporal overlap of the two-color light field is shown ($\Delta t = -100$ fs; scaled to match the intensity). The EDC are obtained by integrating the photoelectron counts over the full photoemission horizon. The dressing of the Au(111) band structure is evident for nearly p-polarized light, whereas in the s-polarized case, no evidence of sidebands is observed. The grey shaded area depicts the integration area used in the analysis in Fig.~4 of the main text for the first ($\mathrm{n_{1}}$, $E-\mathrm{E_{F}}=0.9 - 1.2$~eV) and the zero-order sideband ($\mathrm{n_{0}}$, $E-\mathrm{E_{F}}=-0.3 - 0$~eV). 

Residual photoelectron counts that mainly limit our resolution are induced in linear photoemission by the $13\textsuperscript{th}$ harmonic. These photoelectron counts dominate the measured signal for energies higher than $E-\mathrm{E_{F}}>1.2$~eV, as is evident by the nearly overlapped EDCs in Fig.~\ref{fig:spectrum}. The handling of this background counts in the analysis of the main text is discussed in the next section.

\section{Detailed analysis of the sideband yield in p-polarization}
\label{sec:sidebandyielnormalization}

\subsection{Intensity normalization in Fig.~4~(c)}
\label{sec:normalizationhandling}
\begin{figure*}[hbt!]
    \centering
    \includegraphics[width=\linewidth]{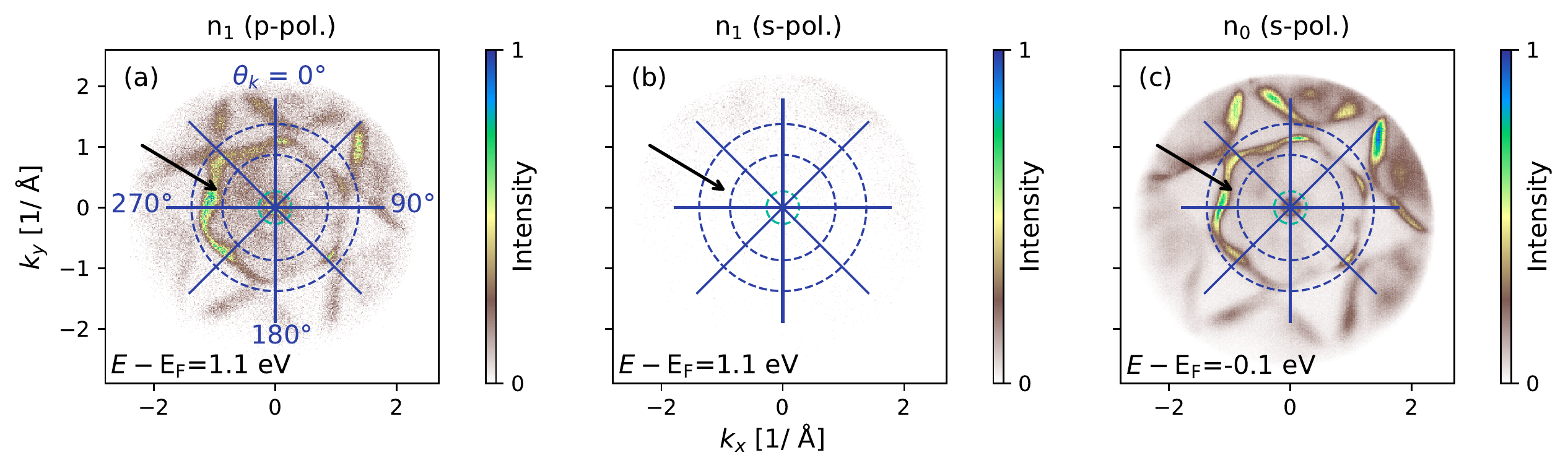}
    \caption{Momentum maps that are used in the analysis in Fig.~4~(c) of the main text. We divide the momentum maps into areas as indicated by the blue lines and integrate all photoelectrons within the region of interest. The reference area is indicated by an arrow. From (a) to (c), we show the first-order sideband (p-polarized driving), the (non-existing) first order sideband (s-polarized driving), and the zero-order sideband (s-polarized driving). The energy of the momentum maps is given in the figure.}  
    \label{fig:data_handling}
\end{figure*}

In Fig.~4~(c) of the main text, the $\theta_k$ dependence of the normalized sideband yield $I_{\mathrm{norm.}}(\theta_k)$ of the SS band and the sp transition are calculated as follows
\begin{equation}
I_{\mathrm{norm.}}(\theta_k)=\frac{\left[I^p_{\mathrm{n_{1}}}(\theta_k)-I^s_{13^{\rm th} }(\theta_k)\right]/I^s_{\mathrm{n_{0}}}(\theta_k)}{\left[I^p_{\mathrm{n_{1}}}(\theta_{k,0})-I^s_{13^{\rm th} }(\theta_{k,0})\right]/I^s_{\mathrm{n_{0}}}(\theta_{k,0})}.
\label{eq:norma}
\end{equation}
In this data handling, we consider the initial observation that s-polarized driving light does not generate sidebands and can thus be used as a reference to normalize the p-polarized case that we are interested in. We divide the momentum maps into areas as shown in Fig.~\ref{fig:data_handling}~(a) and integrate all counts in this area to generate the respective intensity $I(\theta_k)$. First, we consider contributions from the 13th harmonic (see section~\ref{sec:background}) by subtracting $I^s_{13^{\rm th}}$ from the first-order sideband yield $I^p_{\mathrm{n_{1}}}(\theta_k)$. $I^s_{13^{\rm th}}$ is generated by integrating all photoelectron counts in a ($E-\mathrm{E_{F}}=0.9 - 1.2$~eV) interval for the s-polarized measurement [see Fig.~\ref{fig:data_handling}~(b)]. Second, we normalize this difference onto the $\theta_k$-dependent intensity $I^s_{\mathrm{n_{0}}}(\theta_k)$ of the undressed system [see Fig.~\ref{fig:data_handling}~(c)]. This normalization takes care of the initial asymmetry of the undriven momentum distributions due to alignment of the microscope and matrix element effects. The denominator of Eq.~\ref{eq:norma} then references all intensities to the area with highest sideband yield ($\theta_{k,0}$ is the area in Fig.~\ref{fig:data_handling}~(a) labelled with an arrow). $I_{\mathrm{norm.}}(\theta_k)=1$ then implies that the first-order sideband is a perfect replica of the zero-order-sideband.

\subsection{$k_z$-dependent analysis of the sideband yield}

\begin{figure}[hbt!]
    \centering
    \includegraphics[width=0.6\linewidth]{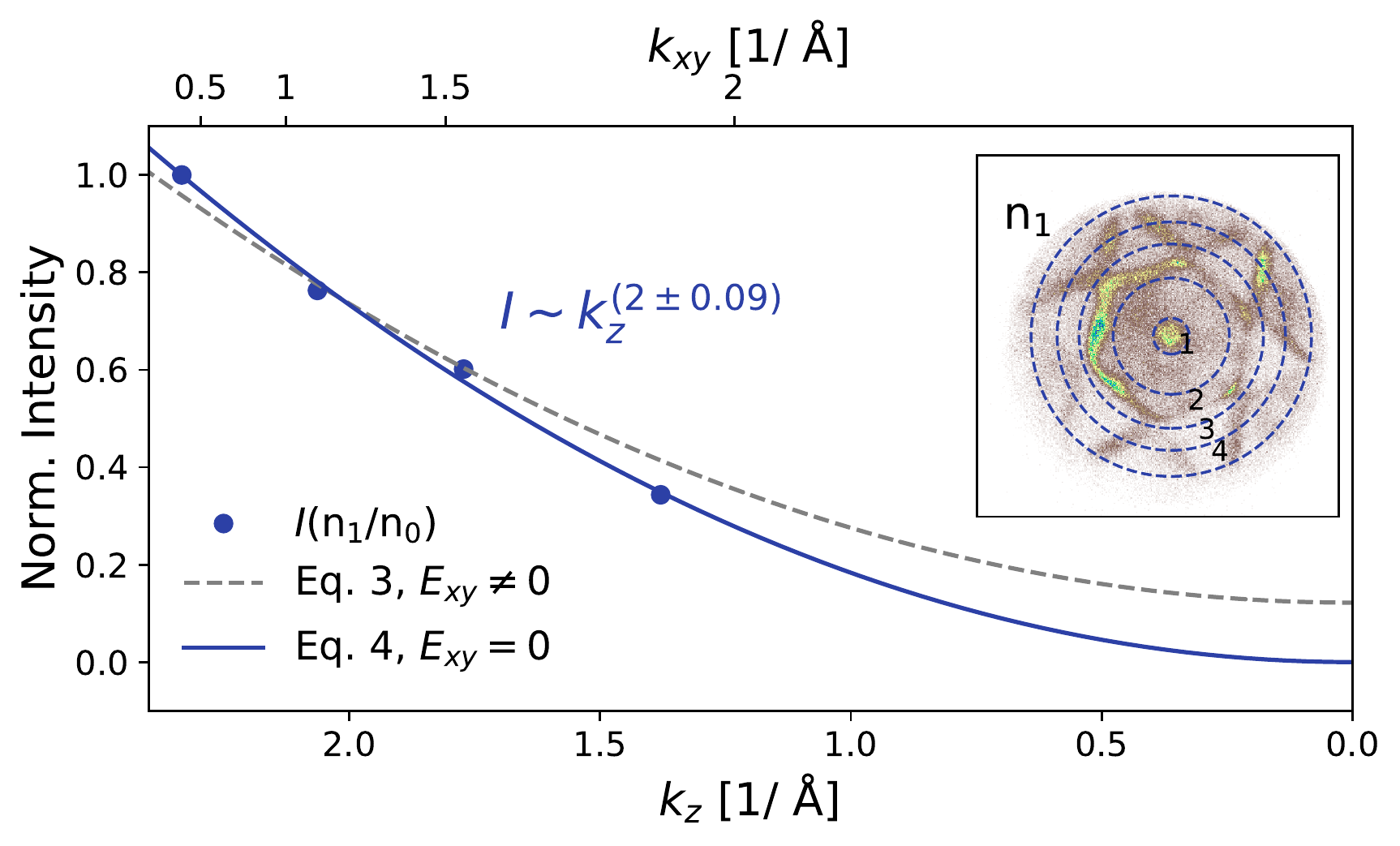}
    \caption{$k_{z}$-dependence of the sideband intensity of the SS band, the sp transition, and the outer part of the second surface Brillouin zone in p-polarized driving, as indicated by the areas 1 to 4 in the inset. When considering screening of the in-plane electric field components, the intensity follows a $k_{z}^2$ behavior as expected from Eq.~\ref{eq:LAPE_screened} (blue line). In contrast, Eq.~\ref{eq:LAPE_not_screened} (without screening) does not describe the data satisfactorily (grey dashed line).}  
    \label{fig:kz}
\end{figure}

As an additional test of our screening model, we check for the $k_{z}$ dependence that is expected from equation \ref{eq:LAPE_screened}. We perform a similar analysis as done in Fig.~4~(c) of the main text and divide the momentum maps into areas 1 to 4 shown in Fig.~\ref{fig:kz} (note that here we integrate over the full $\theta_k$-range). As discussed in section~\ref{sec:normalizationhandling}, we take care of the contribution of the 13\textsuperscript{th} harmonic in the integration process; region 1, that contains photoemission from the surface band is used as a reference. From the value of the in-plane momentum $k_{xy}$ centered in the different areas 1 to 4, we calculate the $k_{z}$ momentum in the free electron picture to
\begin{equation}
k_{z}=\sqrt{\frac{2mE_{kin}}{\hbar^2}-(k_{xy}^2)},
\end{equation}
and plot the normalized intensity as a function of $k_{z}$ momentum (see Fig.~\ref{fig:kz}). As expected from Eq.~\ref{eq:LAPE_screened}, the intensity can be described by a $k_{z}^2$ dependence, which is in accordance to the screening picture.





\section{Calculation of momentum dependent sideband yield}

The theoretical description of the laser-assisted photoelectric effect here follows precedent work by Madsen et al. and others \cite{madsen2005strong,baggesen2008theory,saathoff_laser-assisted_2008,park_interference_2014}. We start with a linearly polarized driving laser field
\begin{equation}
\mathbf{E}(t) = \mathbf{E} \cos(\omega_{IR} t)
\end{equation}
where
\begin{equation}
\mathbf{E}=\begin{pmatrix}E_{xy}\sin(\theta_{E})\\E_{xy}\cos(\theta_{E})\\E_{z}\end{pmatrix}=\begin{pmatrix}E_{0}\cos(68^{\circ})\cos(\phi)\\E_{0}\sin(\phi)\\E_{0}\sin(68^{\circ})\cos(\phi)\end{pmatrix},
\label{eq:LAPE}
\end{equation}
with $\theta_{E}=\arctan\left(\frac{E_{x}}{E_{y}}\right)$, the in-plane angle of the electric field (compare figure 2 in the main text), and $\phi$ the polarization angle,
where $\phi=0$° (90°) corresponds to a p- (s-)polarized driving light.
The latter description can be seen as the transition to the laboratory frame with an incidence angle of $68$°.

\subsection{Model for the final state}

In general, photoemission final states are free electron states in the vacuum that become distorted at the surface and are matched to high-lying Bloch waves giving rise to so-called time-inverted LEED states \cite{PhysRevLett.102.136401}.
For metals a common simple surface model is a step potential model, the so-called "jellium" model \cite{baggesen2008theory}.
Here, the final states are modelled as plane waves parallel to the surface and exponentially damped waves normal to the surface.
The dressing by the laser field, however, can be treated in an approximate way by using the phase of free electron states dressed by a driving laser field \cite{faraggi_interaction_2007,baggesen2008theory}, which are known as Volkov waves \cite{madsen2005strong, wolkow1935klasse}.
Also plain Volkov waves have proven to be useful as final states for analytical photoemission calculations \cite{park_interference_2014}.
Hence, our model for the final state is
\begin{equation}
\phi_{V}(\mathbf{r},t) = \frac{\text{e}^{i \mathbf{k}_{xy} \cdot \mathbf{r}_{xy}}}{2\pi} \phi_{k_z}(z) \text{e}^{-\frac{i}{\hbar} (\hbar\omega_f +U) t} \sum_{n=-\infty}^{\infty}\text{e}^{-i n\omega_{IR} t} J_{n}\left(\alpha,\frac{U}{2\hbar\omega_{IR}}\right),
\label{eq:volkov_wave}
\end{equation} 
where $\phi_{k_z}(z)$ is the surface-normal part of the wave function, $\hbar \omega_f(\mathbf{k})$ is the eigenenergy of the state, which is $\approx \frac{\mathbf{p}^{\prime2}}{2m_e}$. $U=\frac{e^2E_{0}^2}{4m_e \omega_{IR}^2}$ is the ponderomotive potential and $J_{n}$ are generalized Bessel functions.
\begin{equation}
\alpha=\frac{e}{m_e \omega_{IR}^2}\mathbf{E}^\text{IR}\cdot \mathbf{k}
\end{equation}
is the LAPE parameter and $\mathbf{k}$ is the momentum in the final state.

\subsection{Model for the initial state}
The initial state in photoemission is generically a complicated Bloch wave.
However, in the case of the sp band transition the \emph{in-plane} dispersion is well approximated as parabolic
$E_\text{SP}(\mathbf{k}_\text{xy}) = -E_0^\text{SP} + \frac{\hbar^2}{2 m_e} \mathbf{k}_\text{xy}^2$ when measured with respect to $E_\text{Fermi}$.
This is in the spirit of the free electron approximation for noble metals \cite{ashcroft1976solid}, which is a common simple but often quite predictive approximation for many quantities.
Note that $\mathbf{k}_{xy}$ of the inital and final states coincide due to conservation of in-plane momentum in the photoemission process.

A realistic modelling of the perpendicular momentum dispersion is more complicated.
Estimating the initial state $k^\text{in}_z$ from the photoemission energies reveals that for a probing photon energy of 26.5~eV, $k^\text{in}_z$ is typically close to the bulk $\Gamma$-point where the dispersion is flatter than at high momenta, i.e. close to the L point (see band structure calculation and photon energy dependent photoemission data in Ref.~\cite{Rangel12prb} and~\cite{Roth17jesrp}).
For flat dispersions we expect a strong suppression of the sideband generation since the dressing field couples to the momentum-dependent part of the dispersion. This is consistent with the picture of a parabolic dispersion with high effective mass in that direction.

Furthermore, we model the initial state dressing as due to an averaged damped electric field $\mathbf{E}^\text{IR,in}$ inside the metal as a first approximation.
Hence we can work with a wave function similar to Eq.~\eqref{eq:volkov_wave} but consider a Floquet parameter
\begin{equation}
\beta=\frac{e}{m_e \omega_{IR}^2} \mathbf{ E}^\text{IR,in}_{xy} \cdot \mathbf{k}_{xy} + \beta_z(E^\text{IR,in}_z, k^\text{in}_z)
\label{eq:beta_ansatz}
\end{equation}
instead of the LAPE parameter $\alpha$.

The occupied part of the Shockley surface state (SS) has a two-dimensional parabolic dispersion with an effective mass $m_\text{SS}^\ast \approx 0.26 m_e$ \cite{Reinert01prb}:
$E_\text{SS}(\mathbf{k}_{xy}) = -E_0^\text{SS} + \frac{\hbar^2}{2 m_\text{SS}^\ast} {\mathbf{k}_{xy}}^2$.
At the same level of modelling as above we may hence use a Floquet parameter similar to \eqref{eq:beta_ansatz} but with $m_e \rightarrow m_\text{SS}^\ast$ and possibly a different average electric field.


\subsection{Photoemission amplitude}

The transition in photoemission from the initial state to the final state is calculated within first order time-dependent perturbation theory (Born approximation \cite{madsen2005strong,park_interference_2014}) employing a scattering matrix description. The transition amplitude from an initial state $\phi_{i}$ to a final state $\phi_{f}$ reads:
\begin{equation}
(S^{B}-1)_{fi}=-\frac{i}{\hbar}\int_{-\infty}^{\infty}dt\left\langle\phi_{f}\left|\mathbf{A}_\text{EUV}\cdot\hat{p}\right|\phi_{i}\right\rangle,
\label{eq:smatrix_born_approx}
\end{equation}
with the vector potential of the EUV probe $\mathbf{A}_\text{EUV}=\frac{\mathbf{A_{0}}}{2}\exp(-i\omega_\text{EUV}t)$, where we used the dipole and the rotating wave approximation.
In this experiment we have used a driving laser fluence of $F$ = 5 mJ/cm\textsuperscript{2} with $\tau$ = 37 fs pulse duration resulting in an incident electric field amplitude $E_{0}=\sqrt{\frac{2}{c\epsilon_{0}}\frac{F}{\tau}}\approx 1\cdot10^{9}$ V/m.
The ponderomotive potential $U=\frac{e^2E_{0}^2}{4m_{e}w_{IR}^2}$ is then of the order of $10$ meV and the term $\frac{U}{2\hbar\omega_{IR}}\approx 0$ and can therefore be safely neglected. The generalized Bessel function reduces to the ordinary Bessel function of the first kind.


Plugging the model for the sp wave functions into \eqref{eq:smatrix_born_approx} yields
\begin{align} \begin{split}
 (S^\text{B} - 1)_{fi} &= - \frac{i}{\hbar} M_{fi} \sum_{mn} \int_{-\infty}^{\infty} \operatorname{d}{t} \; \text{e}^{i \left( \omega_f(\mathbf{k}) - \omega_i(\mathbf{k}') - (m-n) \omega_\text{IR} - \omega_\text{EUV} \right) t} J_m \big( \beta \big) J_n \big( \alpha \big) \\
 &= - \frac{2\pi i}{\hbar} M_{fi} \sum_{mn} \delta \left( \omega_f(\mathbf{k}) - \omega_i(\mathbf{k}') - (m-n) \omega_\text{IR} - \omega_\text{EUV} \right) J_m \big( \beta \big) J_n \big( \alpha \big) \\
  &= - \frac{2\pi i}{\hbar} M_{fi} \sum_{mn} \delta \left( \omega_f(\mathbf{k}) - \omega_i(\mathbf{k}') - m \omega_\text{IR} - \omega_\text{EUV} \right) J_{n+m} \big( \beta \big) J_n \big( \alpha \big)
 \label{eq:smatrix_born}
\end{split} \end{align}
where
\begin{equation}
 M_{fi} = \Big\langle \phi_f(\mathbf{r} ) \Big| \frac{\mathbf{A}_0^\text{EUV}}{2} \cdot \mathbf{p} \Big| \phi_i(\mathbf{r}) \Big\rangle
\end{equation}
is the photoemission matrix element generated by the spatial parts of the wave function \cite{baggesen2008theory}.
Since we are only interested in the relative $k$-space structure of the sidebands we neglect it in the following.
Using a Bessel function identity allows to simplify the expression for the sideband amplitudes
\begin{align} \begin{split}
 a_m &:= \sum_n J_{n+m} \big( \beta \big) J_n \big( \alpha \big) = J_m \big( \beta - \alpha \big) .
 \label{eq:bessel_sum}
\end{split} \end{align}

The Dirac-$\delta$ in \eqref{eq:smatrix_born} describes energy conservation during photoemission and restricts the final momentum $\mathbf{k}$ in addition to the momentum conservation parallel to the surface.
The photoemission intensity of the $m$-th order sideband is
\begin{equation}
I_m \sim | a_m |^2 = J_m \big( \beta - \alpha \big)^2 = J_m \big( \beta_{xy} - \alpha_{xy} + \beta_z - \alpha_z \big)^2.
\label{eq:sideband_intens_interf}
\end{equation}
Note that the Bessel function obeys: $J_{-m}(\alpha)=(-1)^m J_m(\alpha)$ for integer $m$ so that the intensity of the sidebands $I_m = I_{-m}$.
We can approximate the Bessel function $J_{1}$ for small parameters $| \beta-\alpha | \ll \sqrt{2}$ resulting in a sideband amplitude ($m=1$)
\begin{equation}
|a_1|^2 \sim \frac{1}{4} \left( \beta - \alpha \right)^2.    
\end{equation}
This is justified in our case with an incident electric field $E_{0} = 1\cdot10^{9}$ V/m yielding a maximum LAPE parameter of $\alpha_{\text{max}}=1.28$ and only small corrections due to $\beta$ because of the efficient screening, i.e. $\beta \approx 0$.
Generically, since $\beta_{xy} - \alpha_{xy} = \frac{e}{m_e\omega_{IR}^2} \big( \mathbf{E}_{xy}^\text{IR,in} - \mathbf{E}_{xy}^\text{IR} \big) \cdot  \mathbf{k}_{xy}$
the sideband generation for the sp band states due to the in-plane electric fields will be suppressed if the electric field in the metal is only weakly screened.


\paragraph*{Special case LAPE.}
Since we anticipate strong damping of the electric fields, which is underpinned by a Fresnel equation estimation (cf. next section), we consider the special case $\beta_{xy}, \beta_z \rightarrow 0$ of pure LAPE leading to a sideband amplitude
\begin{equation}
|a_m|^2 \sim J_m 
(\alpha)^2.
\label{eq:LAPE_intensity}
\end{equation}

In previous works \cite{park_interference_2014,mahmood_selective_2016} the in-plane component of the electric field was typically neglected if all electrons are photoemitted nearly perpendicular to the surface ($k_{z} \gg k_{x},k_{y}$).
In our case this approximation is a priori not valid, because we can detect electrons photoemitted under large photoemission angles and thus high in-plane momenta. If the in-plane components of the electric field are taken into account, the result, in contrast, yields a dependence on the azimuthal angle
\begin{equation}
|a_m|^2 \sim J_m \bigg( \frac{e}{m_e\omega_{IR}^2} \Big( E^\text{IR}_{xy} k_{xy} \cos \big( \theta_k - \theta_E \big) + E^\text{IR}_z k_z \Big) \bigg)^2.
\label{eq:sideband_intens_lape}
\end{equation}

We may use the same approximate form of the Bessel function as above 
such that the sideband amplitude for the first sideband ($m=1$) is given by $J_{1}\left(\alpha\right)^2 \approx \frac{\alpha^2}{4}$:
\begin{equation}
|a_m|^2 \sim \frac{1}{4} \bigg( \frac{e}{m_e\omega_{IR}^2} \Big( E^\text{IR}_{xy} k_{xy} \cos \big( \theta_k - \theta_E \big) + E^\text{IR}_z k_z \Big) \bigg)^2
\label{eq:LAPE_not_screened}
\end{equation}

A fit of \eqref{eq:LAPE_not_screened} to the data as shown in Fig. 4 of the main text only works for $E_{xy} = 0$:
\begin{equation}
|a_m|^2 \sim \frac{1}{4} \bigg( \frac{e}{m_e\omega_{IR}^2} E^\text{IR}_z k_z \bigg)^2
\label{eq:LAPE_screened}
\end{equation}

\section{Estimation of the relative contribution of LAPE and Floquet-Bloch states to the sideband yield}

The determination of the relative contribution of Floquet-Bloch and LAPE physics to the overall measured sideband yield is an important question in the verification of Floquet engineering by ARPES.
In general, the spatial dependence of the driving electric fields in comparision to the probability distributions of the involved initial states govern the fraction of the electric field that is able to drive the system. Moreover, the band curvatures and the apparent interference effects need to be considered to fully describe the measured sideband yield by ARPES.
Here we focus on the electric field dependence to show that Floquet-Bloch contributions are strongly suppressed by the efficient screening at the metal surface. Therefore, neither the interference effects nor the inital state dispersion need to be considered to describe the experimental data.

In a simple model we use an abrupt metal-vacuum interface that is described by Frensel equations with the dieletric function for Au \cite{Olmon12prb} ($\epsilon_{1}=-44.252 , \epsilon_{2}=2.0375$).
The relative contribution of Floquet-Bloch bands and sidebands due to the laser-assisted photoeletric effect (LAPE) can then be estimated by comparing the electric fields inside and outside of the metal surface. The orientation of the axis are defined so that the plane of incidence is the xz-plane and the surface of the metal is the xy plane. The electric fields along x- and y-direction, corresponding to an p- and s- polarized incident electric field, can then be written to \cite{whitaker1978use,feibelman1974comment}:
\begin{equation}
\frac{E_{x}^2}{E_{0}^2}=\frac{4\cos(\theta_{d})^2S^{1/2}\epsilon_{d}}{\splitfrac{(\epsilon_{1}^2+\epsilon_{2}^2)\cos(\theta_{d})^2}{+S^{1/2}\epsilon_{d}{\sqrt{2}\cos(\theta_{d})u\epsilon_{d}^{1/2}(S^{1/2}+\epsilon_{d}\sin(\theta_{d})^2})}},
\end{equation}
\begin{equation}
\frac{E_{y}^2}{E_{0}^2}=\frac{4\cos(\theta_{d})^2\epsilon_{d}}{\epsilon_{d}\cos(\theta_{d})^2+S^{1/2}{+\sqrt{2}\cos(\theta_{d})u\epsilon_{d}^{1/2}})}, 
\end{equation}
with the incident electric field amplitude $E_{0}$, the angle of incidence $\theta_{d}$, the vacuum dielectric constant $\epsilon_{d}$ and
\begin{equation}
S=(\epsilon_{1}-\epsilon_{d}\sin(\theta_{d})^2+\epsilon_{2}^2),
\end{equation}
\begin{equation}
u=(S^{1/2}+(\epsilon_{1}-\epsilon_{d}\sin(\theta_{d}))^{1/2}.
\end{equation}
The electric field in z-direction can be calculated below the metal-vacuum interface $E_{z_{-}}$ or above $E_{z_{+}}$:
\begin{equation}
\frac{E_{z_{+}}^2}{E_{0}^2}=\frac{4\sin(\theta_{d})^2\cos(\theta_{d})^2(\epsilon_{1}^2+\epsilon_{2}^2)}{\splitfrac{(\epsilon_{1}^2+\epsilon_{2}^2)\cos(\theta_{d})^2+S^{1/2}\epsilon_{d}}{+{\sqrt{2}\cos(\theta_{d})u\epsilon_{d}^{1/2}(S^{1/2}+\epsilon_{d}\sin(\theta_{d})^2})}},
\end{equation}
\begin{equation}
\frac{E_{z_{-}}^2}{E_{0}^2}=\frac{4\sin(\theta_{d})^2\cos(\theta_{d})^2}{\splitfrac{(\epsilon_{1}^2+\epsilon_{2}^2)\cos(\theta_{d})^2+S^{1/2}\epsilon_{d}}{+{\sqrt{2}\cos(\theta_{d})u\epsilon_{d}^{1/2}(S^{1/2}+\epsilon_{d}\sin(\theta_{d})^2})}}.
\end{equation}
We can then determine that the electric field drops to 2\% inside of the metal $E_{z_{-}}$ in comparison to it's value above the interface $E_{z_{+}}$ using the dielectric function of Au and an angle of incidence of $68$°. Since LAPE is considered to happen at/above the vacuum metal interface, where the surface can act as a source/sink of momentum, the electric field outside of the metallic surface $E_{z_{+}}$ has to be considered. In contrast, in the framework of Floquet-Bloch bands, the optical driving has to interact with the initial states, i.e. the Bloch states, with probability distributions inside of the metal. Therefore, the driving electric field for Floquet-Bloch states is $E_{z_{-}}$. By comparison of $E_{z_{+}}$ and $E_{z_{-}}$ it follows that the observed sideband is dominated by LAPE. This simple estimation shows how critically screening can alter the relative contribution of Floquet-Bloch and LAPE physics to the overall measured sideband yield in an ARPES experiment.  




%